\documentclass[aps,prl,twocolumn,showpacs,superscriptaddress,amsmath,amssymb,nobibnotes, letterpaper]{revtex4-1}

\usepackage{amssymb}
\usepackage[raggedright]{titlesec}
\usepackage{mathtools}
\usepackage[pdftex]{graphicx}
\usepackage[usenames, dvipsnames, svgnames, table]{xcolor}
\usepackage[colorlinks=true, citecolor=Blue,
            linkcolor=BrickRed, urlcolor=ForestGreen]{hyperref}
\usepackage{txfonts}
\usepackage{mathrsfs}
\usepackage{bm}
\usepackage{multirow}
\usepackage{color}
\usepackage{comment}
\usepackage{soul}

\begin{document}

\newcommand{\Com}[1]{{\color{red}{#1}\normalcolor}} 


\title{Demonstration of Maxwell Demon-assisted Einstein-Podolsky-Rosen Steering via Superconducting Quantum Processor}

\author{Z. T. Wang}
\email{These authors contribute equally}
\affiliation{Beijing Academy of Quantum Information Sciences, Beijing 100193, China}

\author{Ruixia Wang}
\email{These authors contribute equally}
\affiliation{Beijing Academy of Quantum Information Sciences, Beijing 100193, China}

\author{Peng Zhao}
\affiliation{Beijing Academy of Quantum Information Sciences, Beijing 100193, China}

\author{Z. H. Yang}
\affiliation{Beijing National Laboratory for Condensed Matter Physics,
Institute of Physics, Chinese Academy of Sciences, Beijing 100190, China}
\affiliation{School of Physical Sciences, University of Chinese Academy of Sciences, Beijing 100190, China}

\author{Yun-Hao Shi}
\affiliation{Beijing National Laboratory for Condensed Matter Physics,
Institute of Physics, Chinese Academy of Sciences, Beijing 100190, China}
\affiliation{School of Physical Sciences, University of Chinese Academy of Sciences, Beijing 100190, China}

\author{Kaixuan Huang}
\email{huangkx@baqis.ac.cn}
\affiliation{Beijing Academy of Quantum Information Sciences, Beijing 100193, China}

\author{Kai Xu}
\affiliation{Beijing National Laboratory for Condensed Matter Physics,
Institute of Physics, Chinese Academy of Sciences, Beijing 100190, China}
\affiliation{Beijing Academy of Quantum Information Sciences, Beijing 100193, China}
\affiliation{CAS Center for Excellence in Topological Quantum Computation, UCAS, Beijing 100190, China}

\author{Yong-Sheng Zhang}%
\email{yshzhang@ustc.edu.cn}
\affiliation{Laboratory of Quantum Information, University of Science and Technology of China, Hefei 230026, China }%
\affiliation{Synergetic Innovation Center of Quantum Information and Quantum Physics,
University of Science and Technology of China, Hefei 230026, China}
\affiliation{Hefei National Laboratory, University of Science and Technology of China, Hefei, 230088, China}

\author{Heng Fan}
\affiliation{Beijing National Laboratory for Condensed Matter Physics,
Institute of Physics, Chinese Academy of Sciences, Beijing 100190, China}
\affiliation{Beijing Academy of Quantum Information Sciences, Beijing 100193, China}
\affiliation{School of Physical Sciences, University of Chinese Academy of Sciences, Beijing 100190, China}
\affiliation{CAS Center for Excellence in Topological Quantum Computation, UCAS, Beijing 100190, China}
\affiliation{Hefei National Laboratory, University of Science and Technology of China, Hefei, 230088, China}
\affiliation{Songshan Lake Materials Laboratory, Dongguan 523808, China}

\author{S. P. Zhao}
\affiliation{Beijing National Laboratory for Condensed Matter Physics,
Institute of Physics, Chinese Academy of Sciences, Beijing 100190, China}
\affiliation{School of Physical Sciences, University of Chinese Academy of Sciences, Beijing 100190, China}
\affiliation{CAS Center for Excellence in Topological Quantum Computation, UCAS, Beijing 100190, China}
\affiliation{Songshan Lake Materials Laboratory, Dongguan 523808, China}


\author{Meng-Jun Hu}
\email{humj@baqis.ac.cn}
\affiliation{Beijing Academy of Quantum Information Sciences, Beijing 100193, China}

\author{Haifeng Yu}
\affiliation{Beijing Academy of Quantum Information Sciences, Beijing 100193, China}
\affiliation{Hefei National Laboratory, University of Science and Technology of China, Hefei, 230088, China}
\date{\today}

\begin{abstract}
The concept of Maxwell demon plays an essential role in connecting thermodynamics and information theory, while entanglement and non-locality are fundamental features of quantum theory. 
Given the rapid advancements in the field of quantum information science, there is a growing interest and significance in investigating the connection between Maxwell demon and quantum correlation.
The majority of research endeavors thus far have been directed towards the extraction of work from quantum correlation through the utilization of Maxwell demon.
Recently, a novel concept called Maxwell demon-assisted Einstein-Podolsky-Rosen (EPR) steering has been proposed, which suggests that it is possible to simulate quantum correlation by doing work. This seemingly counterintuitive conclusion is attributed to the fact that Alice and Bob need classical communication during EPR steering task, a requirement that does not apply in the Bell test. 
In this study, we demonstrate Maxwell demon-assisted EPR steering with superconducting quantum circuits. 
By compiling and optimizing a quantum circuit to be implemented on a 2D superconducting chip, we were able to achieve a steering parameter of $S_{2}=0.770\pm 0.005$ in the case of two measurement settings, which surpasses the classical bound of $1/\sqrt{2}$ by $12.6$ standard deviations.
In addition, experimental observations have revealed a linear correlation between the non-locality demonstrated in EPR steering and the work done by the demon. Considering the errors in practical operation, the experimental results are highly consistent with theoretical predictions.
Our findings not only suggest the presence of a Maxwell demon loophole in the EPR steering, but also contribute to a deeper comprehension of the interplay between quantum correlation, information theory, and thermodynamics.
\end{abstract}

\maketitle
\noindent{\textbf{INTRODUCTION}}\\
Non-locality is a key characteristic of quantum theory, which sets it apart from traditional theories that adhere to local realism \cite{EPR1, EPR2, EPR3}. In his pioneering paper, Bell shows that quantum theory refutes any local hidden variable models \cite{Bell}. Since then, great efforts have been made to verify this Bell non-locality with loopholes closed using the Bell-CHSH inequality \cite{CHSH1, CHSH2,CHSH3,CHSH4, CHSH5, CHSH6, CHSH7, CHSH8, CHSH9, CHSH10, CHSH11, CHSH12}. Non-locality is attributed to quantum correlation, and the existence of its hierarchy has been shown, especially when Einstein-Podolsky-Rosen (EPR) steering was discovered by Wiseman {\it et al} \cite{steering1}. EPR steering, which refutes any local hidden state models, is a form of non-locality that lies between Bell non-locality and non-separability \cite{steering2}. 

From an information theory standpoint, it can be observed that the entropy of an entangled quantum system is comparatively lower than that of mixed systems after measurements. The observation suggests that the entangled system contains additional information that can be harnessed to generate useful work \cite{work}. 
Since the concept of the Maxwell demon is deeply connected to both thermodynamics and information \cite{demon}, it is natural to consider the interplay between the Maxwell demon and quantum correlations. 
Zurek initially analyzed Szilard's one-molecule engine in the quantum domain \cite{Zurek1}. Subsequently, numerous studies have been conducted on the detection of quantum correlation through the utilization of Maxwell's demon \cite{Zurek2, entanglement1, entanglement2, entanglement3, entanglement4, entanglement5}. 
In recent years, there has been a significant surge of interest in the quantum heat engine (QHE) \cite{QHE1, QHE2, QHE3, QHE4, QHE5} that incorporates the Maxwell demon. Whether quantum resources, such as quantum correlation, can demonstrate the quantum advantage of QHE has been a central topic of investigation \cite{advantage1, advantage2, advantage3}. Beyer {\it et al} showed that quantum steering can be applied to verify the quantumness of Szilard's engines, in which the violation of a steering inequality is connected to the macroscopic average work that outputs the classically extractable work\cite{Beyer}. Du {\it et al} experimentally demonstrated the quantum advantage of this quantum Szilard's engines in the diamond when the demon can  ``steer" the working medium where a steering inequality is violated \cite{Du}.

\begin{figure*}[tbp]
\centering
\includegraphics[width=0.9\textwidth]{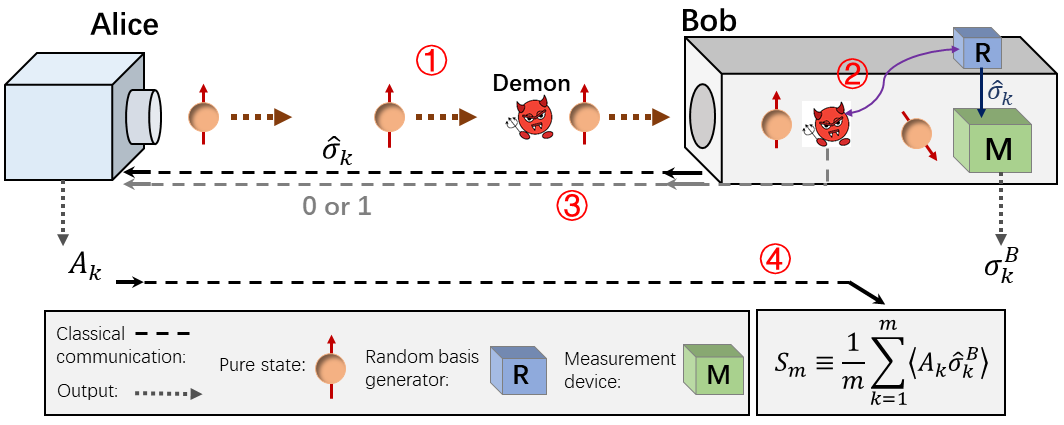}
\caption{
The Maxwell Demon-assisted EPR steering task. { (1)} Bob receives qubits sent by Alice, meanwhile the demon sneaks into Bob's device along with qubits. All qubits are prepared in the same pure state. {(2)} The demon gains access to the information of the measurement setting $\hat{\sigma}_{k}$ by entangling itself with Bob's random basis generator. The demon then randomly rotates the qubits into one of the eigenstates of $\hat{\sigma}_{k}$ before Bob performs measurements. {(3)} Bob announces his measurement setting $\hat{\sigma}_{k}$ to Alice, meanwhile the demon secretly sends an additional $1$ bit of information to Alice, which is mixed in the communication channel between Bob and Alice. The $1$ bit information tells Alice which eigenstate of $\hat{\sigma}_{k}$ it has chosen to rotate the qubits into. {(4)} Based on the information provided by Bob and the demon, Alice responds to Bob regarding her result $A_{k}$. {(1)-(4)} are repeated multiple times, and Bob combines the results to calculate the steering parameter $S_{m}$. If $S_{m}$ exceeds a certain classical bound, then Bob becomes convinced that Alice has the ability to remotely manipulate his state. It should be noted that the demon may not do operations for every run, and in this case, Alice answers Bob randomly.
}
\label{f1}
\end{figure*} 

To date, researchers have primarily focused on extracting work by leveraging quantum correlation \cite{attention1, attention2, attention3}. It is intriguing and significant to contemplate alternative viewpoints and inquire about the possibility of simulating quantum correlation or non-locality through work. The motivation behind this is twofold. Firstly, since quantum correlation can be used to extract work, it is natural to ask whether doing work can simulate quantum correlation from an academic point of view. Secondly, the possibility of simulating quantum correlation through work implies the existence of possible thermodynamic loopholes in the non-locality tests that have not been considered before. In a recent study, the authors (including two of the present authors) provided a positive response to the question posed \cite{Hu}. The incorporation of a Maxwell demon into the EPR steering task, i.e., Maxwell demon-assisted EPR steering has been proposed, enabling collaboration between Alice and a demon to deceive Bob through local operations and classical communication. The reason for this discrepancy lies in the presence of information exchange between Alice and Bob during the task of EPR steering, a phenomenon that is not possible in the context of the Bell test. The concept of the Maxwell demon-assisted EPR steering introduces a new loophole in thermodynamics, which is distinct from existing loopholes related to locality \cite{locality1, locality2, locality3}, free will \cite{free} and detection \cite{detection}. The closure of this demon loophole requires continuous monitoring of Bob's surroundings to ensure that there are no abnormal heat fluctuations caused by the demon's attempts to erase its memory. In order to investigate the correlation between the work performed by the demon and the non-locality exhibited in the EPR steering, a dedicated quantum circuit model has been proposed \cite{Hu}.
The novelty of Maxwell demon-assisted EPR steering lies in demonstrating that the non-locality demonstrated in EPR steering can be simulated by the demon's work through local operations and classical communications, which proves the existence of thermodynamic loophole in the non-locality verification beyond the Bell test.

\begin{figure*}[!htbp]
\centering
\includegraphics[width=1\textwidth]{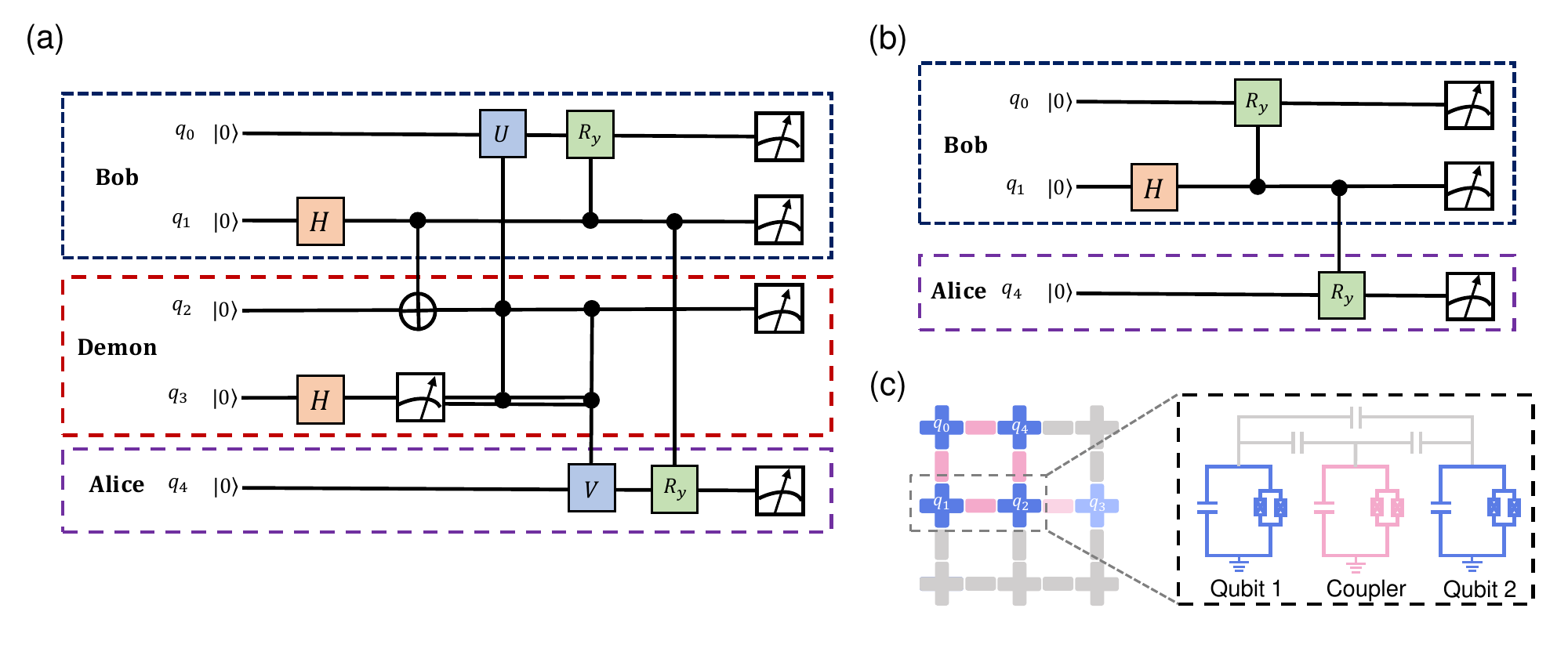}
\caption{{ (a)} Quantum circuit representation of Maxwell demon-assisted EPR steering with two measurement settings $\lbrace\hat{\sigma}_{Z}, \hat{\sigma}_{X}\rbrace$. { (b)} Quantum circuit representation without the Maxwell demon. { (c)} Schematic diagram of a superconducting quantum processor in which the physical qubits $\rm q_{0}-q_{4}$ correspond to the logical qubits $\rm q_{0}-q_{4}$ in the quantum circuit.
In the practical experiment, we execute circuit (a) or circuit (b) with a certain probability $p$ to simulate the case that the demon performs operations probabilistically. }
\label{f2}
\end{figure*}

In this work, we experimentally demonstrate the Maxwell demon-assisted EPR steering with superconducting qubits as the demon \cite{Lloyd}. The measurement and feedback of the demon are implemented via appropriate gate operations. The process of erasing a demon's memory involves resetting qubits to their ground state, which results in the emission of microwave photons into the local environment. By compiling and optimizing a demon circuit to accommodate the physical qubits on a 9-qubit superconducting chip, we have observed quantum correlation between Alice and Bob, where the EPR steering parameter exceeds the classical bound. More significantly, we have successfully validated the linear correlation between the non-locality exhibited in EPR steering and the work performed by the demon. In addition, we provide detailed protocols on realizing locality loophole-free Maxwell demon-assisted EPR steering and Maxwell demon loophole-free EPR steering. Our work shows that superconducting qubits are an excellent platform for studying the relationship between quantum correlations, thermodynamics, and information theory.

\noindent{\textbf{RESULTS}}\\
\noindent{\textbf{Maxwell demon-assisted EPR steering}}\\
The illustration of Maxwell demon-assisted EPR steering task is presented in Fig.~\ref{f1}.  
The task consists of four stages. In the first stage, Alice sends the qubits that are prepared in the same state $|0\rangle$ to Bob. Meanwhile, the demon sneaks into Bob's measurement device accompanying the qubits. In the second stage, the demon first gets access to the information of the measurement setting $\hat{\sigma}_{k}$ by entangling itself with the random basis generator of Bob. Suppose that the standard state of the demon is denoted as $|D=0\rangle$. The entangled state can be expressed as $\sum_{k=1}^{m}|D=k\rangle\otimes|k\rangle$, where the orthogonal states $\lbrace |k\rangle\rbrace$ of the generator represent Bob's different measurement settings choices $\lbrace\hat{\sigma}_{k}\rbrace$. 
After selecting one of the measurement settings $\hat{\sigma}_{k}$, and before the qubit measurement, there exists a temporal window during which the demon can manipulate the qubit through operations. Since the demon knows the measurement setting $\hat{\sigma}_{k}$, it can randomly rotate the qubit into one of the eigenstates of $\hat{\sigma}_{k}$ before Bob performs measurement. In the third stage, Bob announces his measurement setting $\hat{\sigma}_{k}$ to Alice through a classical communication channel. During the communication between Alice and Bob, the demon can secretly send $1$ bit of additional information mixed in the channel that tells Alice which eigenstates of $\hat{\sigma}_{k}$ it has chosen to rotate the qubit into. In the last stage, Alice gives back her result $A_{k}$ to Bob based on the information provided by Bob and the demon. The above four stages are repeated multiple times during the whole task. At last, Bob combines all the results to calculate the steering parameter $S_{m}$ to check if Alice has the true ability to remotely steer his qubit state. To reduce the risk of being detected the demon may not do operations for every run, and in this case, Alice answers Bob randomly. The presence of the above protocol can be attributed to the necessity for Bob and Alice to engage in classical communication during the steering task, a requirement that does not apply to the Bell test.

\noindent{\textbf{Quantum circuit realization}}\\
Currently, the majority of non-locality tests are conducted using photonic systems \cite{test1, test2, test3}. However, it appears to be challenging to demonstrate Maxwell demon-assisted EPR steering, as depicted in Fig.~\ref{f1}. Fortunately, the corresponding quantum circuit model provides a promising way to demonstrate it on current quantum processors \cite{NISQ}. Fig.~\ref{f2}(a) illustrates a specific quantum circuit representation of Maxwell demon-assisted EPR steering with two measurement settings $\lbrace\hat{\sigma}_{Z}, \hat{\sigma}_{X}\rbrace$. The demon is composed of two qubits, in which one qubit $q_{2}$ is entangled with Bob's basis generator $q_{1}$ via a $\mathrm{CNOT}$ gate, and the other $q_{3}$ is used to determine the rotation of qubit $q_{0}$ to be measured. The crucial component of the circuit is the three qubit control-control-$U$ gate, which describes how the demon does an operation on the obtained measurement setting information. Specifically, the operation $U$ is determined as \cite{Hu}
\begin{equation}
U|b\rangle_{q_{3}}|a\rangle_{q_{2}}|0\rangle_{q_{0}}=|b\rangle_{q_{3}}|a\rangle_{q_{2}}\otimes e^{-i\frac{\pi}{4}(a+2b)\hat{\sigma}_{Y}}|0\rangle_{q_{0}}
\end{equation}
with $a, b\in\lbrace 0, 1\rbrace$. In the given context, the variable $a$ is used to represent different measurement settings. Specifically, when $a=0$, it corresponds to the measurement setting $\hat{\sigma}_{Z}$, while $a=1$ represents the measurement setting $\hat{\sigma}_{X}$. On the other hand, the variable $b$ takes the values of either 0 or 1 corresponding to the +1 or -1 eigenstate of the chosen setting, respectively. For instance, when $a=1$ and $b=0$, it follows that the measurement setting $\hat{\sigma}_{X}$ is selected, and the qubit $q_{0}$ is rotated into its $+1$ eigenstate $|+\rangle=(|0\rangle+|1\rangle)/\sqrt{2}$. The circuit incorporates the control-control-V gate and the control-$R_{y}$ gate, which together simulate the process of Alice responding to Bob using information obtained from the demon and Bob. The operators $V$ and $U$ are subject to the condition $\langle 0|V^{\dagger}U|0\rangle=1$ in order to replicate a positive correlation in our experiment. In each run, the demon has to erase its memory to the standard state $|D=0\rangle$, which is equivalent to resetting the qubits in the circuit. The process of memory erasure, as described by Landauer's erasure principle \cite{erasure}, results in the dissipation of energy into the local environment. Specifically, this energy is emitted as microwave photons within the circuit case. In order to mitigate the likelihood of detection by Bob, the demon should perform operations probabilistically during each run. For the scenario where the demon remains inactive, Alice responds to Bob in a random manner. The corresponding quantum circuit is illustrated in Fig.~\ref{f2}(b). Suppose that the demon does operations with a probability of $p$ for each run. The resulting steering parameter $S_{2}$ is determined by \cite{Hu}.
\begin{equation}
S_{2} = \frac{1}{2}(1+p).
\end{equation}
When $p>\sqrt{2}-1$, EPR steering is demonstrated. Given that the reset process of qubits remains consistent across multiple experimental runs, the equation presented above illustrates the linear correlation between the non-locality demonstrated in EPR steering and the work performed by the demon.

\begin{figure}[tbp]
\centering
\includegraphics[width=0.45\textwidth]{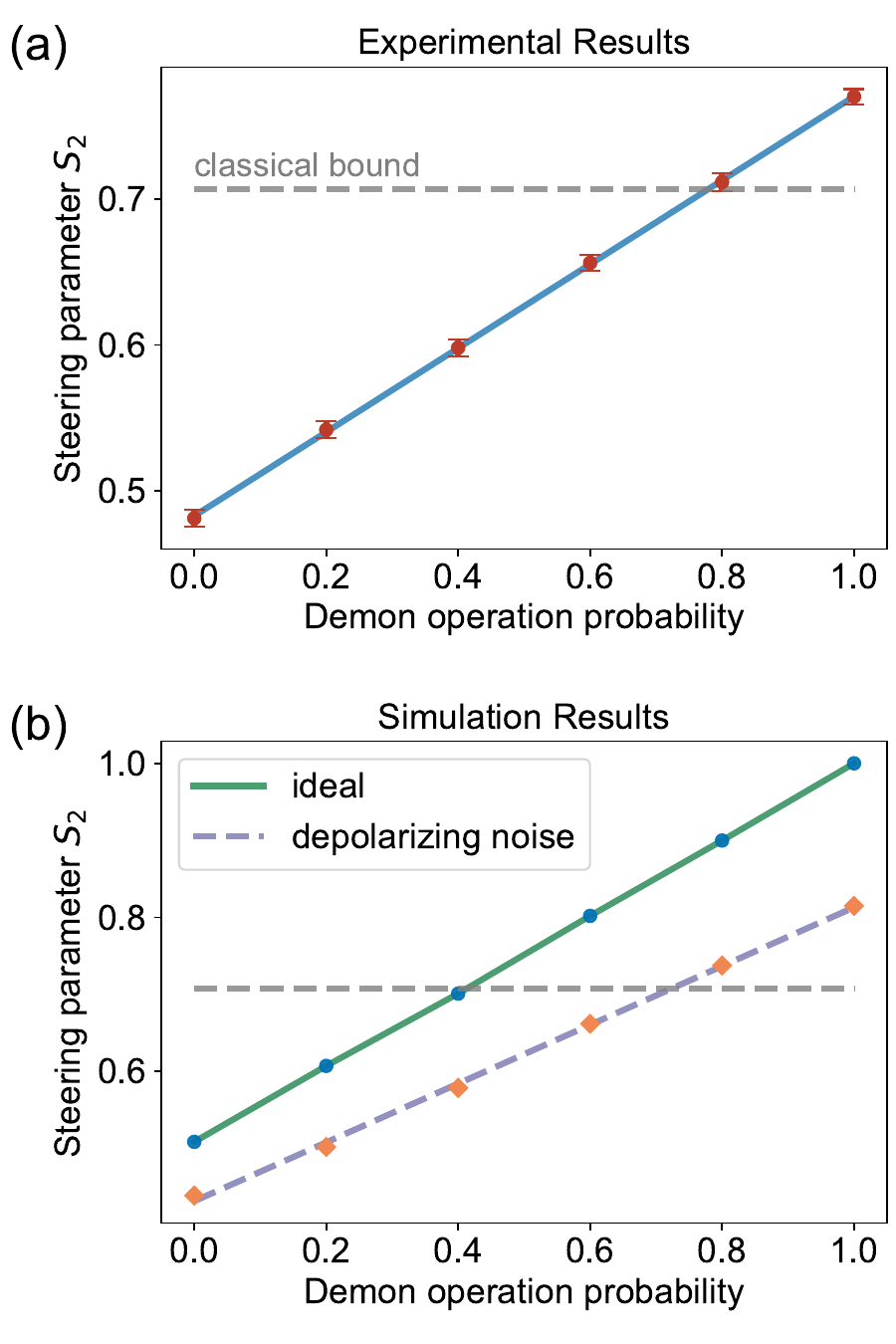}
\caption{
(a) Experimental results verify the linear relationship between the probability of demon operation $p$ and the steering parameter $S_{2}$, as predicted by theory. The largest value of $S_{2}$ is measured to be $0.770\pm 0.005$, which exceeds the classical bound of $1/\sqrt{2}$ by $12.6$ standard deviations. (b) Simulation results show that the deviation between the experiment and the ideal case is primarily caused by depolarizing noise in the actual circuits.
}
\label{data}
\end{figure}

\noindent{\textbf{Experimental results}}\\
The experiment is performed on a transmon-type two-dimensional square grid superconducting processor, as depicted in Fig.~\ref{f2}(c). The processor consists of 9 qubits and 12 couplers, with five qubits $\rm q_{0}-q_{4}$ being utilized for executing the quantum circuits. The fabrication of the superconducting chip involves the utilization of flip-chip technology, incorporating a tantalum base layer \cite{device}. Each qubit in the device is in Xmon form and controlled by the control line combining microwave drive (XY control) and flux bias (Z control).
The quantum state of each qubit is achieved through capacitive coupling to a quarter-wave resonator, enabling the simultaneous achievement of states for all qubits. The nearest neighbor qubits are connected by the coupler which is also a transmon qubit. The effective coupling strength of neighboring qubits can be continuously adjusted from $\sim +2\mathrm{MHz}$ to $\sim -50\mathrm{MHz}$ by manipulating the frequency of the coupler using an individual flux bias line. The energy relaxation time $T_{1}$ for each qubit at the idle point is around $30\,\mu$s, and the dephasing time $T_{2}^{*}$ varies from $1.60\,\mu$s to $4.23\,\mu$s. One triggering period lasts for 310 $\mu$s, and the remaining duration after the operation allows ample time for the qubits to reset to the state $|0\rangle$. In our experiment, all single qubit gates are implemented by using cosine enveloped pulses, and we employ the derivative removal by adiabatic gate (DRAG) technique to effectively minimize leakage \cite{DRAG}. The non-adiabatic CZ gate is implemented by adjusting the energy levels gradually with an error function-shaped pulse. We optimize the square pulses of the qubits and the couplers in order to mitigate leakage and attain the intended conditional phases of the CZ gate. By using Randomized Benchmarking (RB) \cite{RB}, we are able to attain an average fidelity of $99.9\%$ for single qubit gate and $99\%$ for the two-qubit CZ gate, respectively (see supplementary information I-IV for more details).

With high-fidelity single-qubit gates and two-qubit CZ gates as the elementary gates, the quantum circuit of Maxwell demon-assisted EPR steering can be decomposed into the sequence of elementary gates. In order to ensure accurate results, we map the logical qubits onto the physical qubits of the quantum device in such a way that the depth of the executed quantum circuits is as shallow as possible. The most important operation in the circuit is the three qubits control-control-$U$ gate described by Eq. (1). Since the quantum operation between $q_{0}$ and $q_{2}$ depends on the measurement results of $q_{3}$, which acts as a random choice generator for the demon, we can divide the original circuit into two equivalent sub-circuits. Each sub-circuit consists of only two qubit operations, based on the two possible results of $q_{3}$. The two sub-circuits can be further compiled and optimized into elementary gates based on the structure of the quantum device (see supplementary information V for more details). In the Maxwell demon-assisted EPR steering experiment, we first perform a measurement on $q_{3}$ and then execute the corresponding sub-circuit based on its measurement result.

In this experiment, we execute the quantum circuit of Maxwell demon-assisted EPR steering with probability $p$ varying from $0$ to $1$ to simulate the case that the demon performs operations probabilistically. The experimental results are given in Fig.~3(a), illustrating a distinct linear correlation between the steering parameter $S_{2}$ and the operational probability of the Maxwell demon. Given that the reset process of the qubits remains consistent across multiple runs, it can be inferred that the dissipated energy of microwave photons resulting from the demon's work is the same in each run. The obtained results confirm the linear correlation between non-locality, as demonstrated in EPR steering, and the work done by the demon. The largest value of $S_{2}$ is $0.770\pm 0.005$, which exceeds the classical bound of $1/\sqrt{2}$ by $12.6$ standard deviations. We conducted quantum state tomography to examine the simulated Bell state of Alice-Bob in the largest $S_{2}$ case and obtained an average fidelity of 0.951 compared to the ideal Bell state. Due to the imperfection of the current quantum processor \cite{Sup1, Sup2, Sup3}, there exists a deviation between the experimental results and the ideal one described by Eq. (2). The simulation of compiled noisy quantum circuits demonstrates that depolarizing is the primary contributor of noise, as depicted in Fig.~\ref{data}(b).

\begin{figure}[tbp]
\centering
\includegraphics[width=0.48\textwidth]{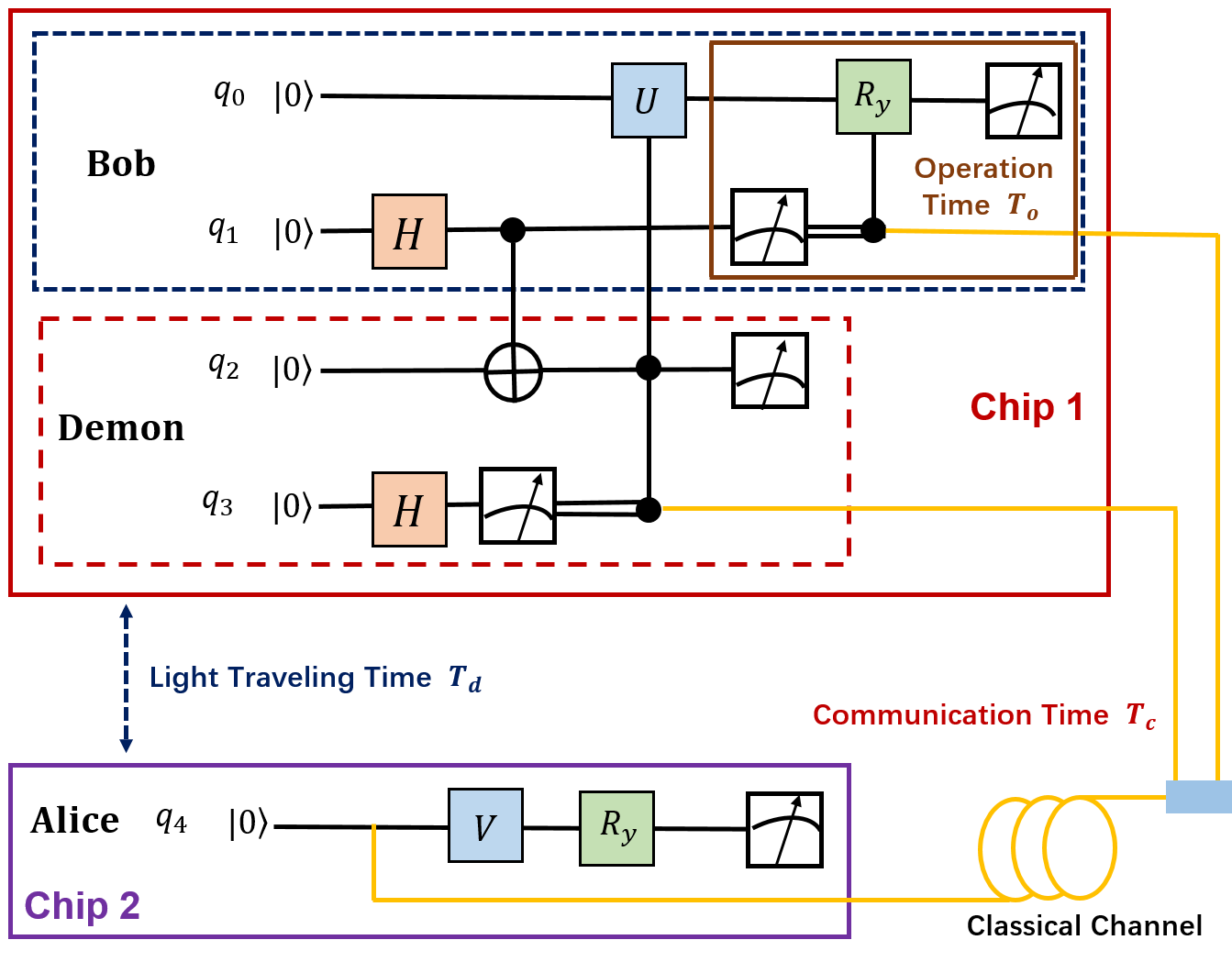}
\caption{Illustration of locality loophole-free Maxwell demon-assisted EPR steering. Alice and Bob are physically separated such that Bob's measurement is not causally influenced by any potential physical signals from Alice, which requires $T_{o} < T_{d}$ in practice.} 

\label{newP}
\end{figure}

\noindent{\textbf{DISCUSSION}}\\
\noindent{\textbf{Maxwell demon-assisted EPR steering without the locality loophole}}\\
We have demonstrated the Maxwell demon-assisted EPR steering via superconducting quantum circuits, implying the existence of a thermodynamic loophole in the task of EPR steering. It would be natural to ask further how to realize Maxwell demon-assisted EPR steering without the locality loophole. With the close of the locality loophole, we will truly prove that the non-locality demonstrated in the EPR steering can be simulated through the work of the demon with the help of local operation and classical communication. The proposal for experimental realization of locality loophole-free Maxwell demon-assisted EPR steering has been illustrated in Fig.~\ref{newP}. Compared to Fig.~\ref{f2}(a), the qubit $q_{4}$ that holds in Alice should be physically separated from Bob. In practice, we can put two quantum chips in two distant locations with light traveling time $T_{d}$. Interestingly, the two quantum chips can have the same or different types of physical qubits, e.g., chip $1$ consists of superconducting qubits while chip $2$ is made of an ion qubit. The closeness of the locality loophole requires that the measurement of Bob's qubit $q_{0}$ will not be causally influenced by any potential physical signals from Alice. Specifically, it requires that the operation time $T_{o}$ is less than the light traveling time $T_{d}$, i.e., $T_{o} < T_{d}$. 
In addition, there exists communication time $T_{c}$ which represents the information exchange time between Alice and Bob through the classical channel. It is obvious that $T_{o} < T_{c}$ in the locality loophole-free setting.

\noindent{\textbf{Maxwell demon loophole-free EPR steering}}--
Demon loophole in the EPR steering is a previously unrecognized thermodynamic loophole. To close the demon loophole, the thermodynamic effects have to be considered. The physical size of the demon must be small enough so that it sneaks into Bob's device without being detected. The small physical size implies that the memory capacity of the demon is highly restricted. The demon has to erasure its memory to the standard state $|D\rangle$ after each task run as illustrated in Fig.~\ref{f1}. The erasure of the memory corresponds to the reset of demon qubits, i.e., $q_{2}, q_{3}$ to the ground state in the quantum circuit as shown in Fig.~\ref{f2}a. The erasure of the memory is a logically irreversible process, there is at least $\mathrm{kTln2}$ heat energy dissipated into the local environment of erasing $1$ bit of information according to Landauer's erasure principle \cite{erasure}. The dissipated heat energy corresponds to microwave photons emitted by the demon qubits due to the reset process since the superconducting qubit operates on several $\mathrm{GHz}$. If the experiment runs enough times the emitted microwave photons can be detected by the most advanced microwave photon detector. To realize Maxwell demon-loophole EPR steering, Bob can continuously monitor his local environment to check if additional microwave photons exist for enough task runs. If no additional microwave photons are detected during enough task runs, then Bob can be confident that Alice is not cheating on him with the demon's help. This would close the demon loophole in the practical EPR steering task.

\noindent{\textbf{CONCLUSION}}\\
In conclusion, we have experimentally demonstrated the Maxwell demon-assisted EPR steering via the superconducting quantum processor. Our results reveal that quantum correlation can be effectively simulated through the application of work. Furthermore, we have identified the presence of Maxwell demon loophole in the EPR steering. 
It should be emphasized that in the case of the Bell test, the Maxwell demon loophole does not exist, as classical communication between Alice and Bob is prohibited during the task. 
With current state-of-the-art technologies in superconducting chips \cite{new}, we can potentially achieve the Maxwell demon-assisted EPR steering task without the locality loophole. By using the most advanced microwave photon detection technology, it is also possible to realize Maxwell demon loophole-free EPR steering. The Maxwell demon-assisted EPR steering demonstrated provides a novel approach to further explore the interplay between quantum correlation, information theory and thermodynamics.

\hfill

\noindent{\textbf{Acknowledgments}}\\
This work was supported by the National Natural Science Foundation of China (Grants Nos. 92365206, 920651131, T2121001, and T2322030), the Innovation Program for Quantum Science and Technology (Grant No.
2021ZD0301802), and the Beijing Nova Program (Grant No. 2022000216).

\noindent{\textbf{Author contributions}}\\
H. Yu, S. Zhao, H. Fan and M. Hu start the project.  M. Hu and Y. Zhang complete the experimental proposal and theoretical analysis.
Z. Wang performed the experiment and collected the experimental results.  R. Wang ran the numerical simulations. Z. Wang and Z. Yang contributed to quantum system calibration. Z. Wang, Z. Yang, Y. Shi, K. Huang, and K. Xu contributed to quantum system stability improvements. All authors contributed to writing the manuscript.

\noindent\textbf{Competing interests}\\
The authors declare no competing interests.

\end{document}